\documentstyle[12pt]{article}

\newcommand{\f}[1]{\,\mbox{\raisebox{0.15ex}{\footnotesize $#1$}}\,}

\begin{document}



\thispagestyle{empty}
 \begin{flushright}
 {hep-ph/9604270}\\
 {April 1996}     \\
\end{flushright}
 \vspace*{3cm}
\begin{center}
 {\bf \large
The sunset diagram in SU(3) chiral perturbation theory}
\vspace{10 mm} \\
  P.~Post and J.B.~Tausk
  \footnote{E-mail: post@vipmzw.physik.uni-mainz.de,
                   tausk@vipmzw.physik.uni-mainz.de}
\vspace{10 mm} \\
{\em
  Institut f\"ur Physik,
  Johannes Gutenberg Universit\"at,\\
  Staudinger Weg 7, D-55099 Mainz, Germany}
\end{center}
\vspace{12 mm}
\begin{abstract}
A general procedure for the calculation of a class of two-loop Feynman
diagrams is described. These are two-point functions containing three
massive propagators, raised to integer powers, in the denominator, and
arbitrary polynomials of the loop momenta in the numerator. The ultraviolet
divergent parts are calculated analytically, while the remaining finite
parts are obtained by a one-dimensional numerical integration, both below
and above the threshold. Integrals of this type occur, for example, in
chiral perturbation theory at order $p^6$. 

\end{abstract}
\newpage
\setcounter{page}{2}
\setcounter{footnote}{0}

\newpage


\section{Introduction}

During the last couple of years there have been growing efforts to extend
the predictions of chiral perturbation theory, which is an effective theory
of QCD at low energies, to the order ${\cal O}(p^6)$ of the momentum
expansion. Of special interest are processes which vanish at
${\cal O}(p^2)$: examples such as $\gamma\gamma \f{\to} \pi^0\pi^0$
\cite{JGasser} or $\eta \f{\to} \pi^0\gamma\gamma$ \cite{Jetter} show that
the ${\cal O}(p^6)$ contributions can be rather large but necessary to find
agreement between the theoretical predictions and experimental data.
In going beyond ${\cal O}(p^4)$ one is confronted with two major
difficulties. On the one hand, the chirally invariant structures of order
${\cal O}(p^6)$ \cite{SScherer} come with a number of a priori unknown
constants which are to be determined from experiment or estimated in some
phenomenological model. On the other hand, at order ${\cal O}(p^6)$, the
diagrammatic expansion involves two-loop Feynman integrals. In our paper we
focus on the second point and discuss a class of two-loop Feynman diagrams
which arise naturally in the context of SU(3) chiral perturbation theory,
namely integrals of the {\em sunset}\/ topology for {\em arbitrary internal
masses, tensor numerators}\/ and (for reasons which will become clear in a
moment) {\em powers of denominators}. The necessity to consider the
arbitrary mass case is a characteristic distinction between SU(2) and SU(3)
chiral perturbation theory.
It is due to the presence (in the limit of isospin conservation) of
three different masses ($m_\pi$, $m_K$, $m_\eta$) in the latter, as
opposed to only one in the former ($m_\pi$). 
As a consequence, the calculation of the genuine two-loop diagrams becomes
essentially more complex.

Any field theory containing four-point interaction vertices with derivative
couplings produces sunset-type integrals with numerators at the two-loop 
level. This is obvious for self-energies, but is also the case,
for example, if one studies form factors for small momentum transfer, such
as electromagnetic charge-radii of mesons. A typical contribution is
represented by the following diagram (the wiggly line is a photon, the
others are mesons):  
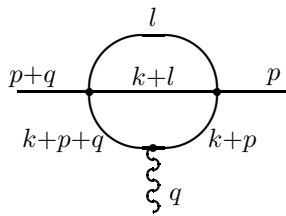
\begin{equation}
\label{fig3pt}
\unitlength1mm
\begin{picture}(50,30)
\thicklines
\put(30,20){\oval(17,15)}         
\put(30,20){\line(1,0){18}}       
\put(12,20){\line(1,0){18}}       
\put(21.5,20){\circle*{1}}
\put(38.5,20){\circle*{1}}
\put(30,12.5){\circle*{1}}
\put(30,11.75){\oval(1.5,1.5)[l]}  
\put(30,10.35){\oval(1.5,1.5)[r]}
\put(30,8.95){\oval(1.5,1.5)[l]}  
\put(30,7.55){\oval(1.5,1.5)[r]}
\put(30,6.15){\oval(1.5,1.5)[l]}
\put(30,4.75){\oval(1.5,1.5)[r]}  
\put(14,22){\makebox(0,0){\footnotesize $p$+$q$}}   
\put(46,22){\makebox(0,0){\footnotesize $p$}}
\put(30,30){\makebox(0,0){\footnotesize $l$}}
\put(18,13.5){\makebox(0,0){\footnotesize $k$+$p$+$q$}}
\put(40.5,13.5){\makebox(0,0){\footnotesize $k$+$p$}}
\put(33,6){\makebox(0,0){\footnotesize $q$}}
\put(30,22){\makebox(0,0){\footnotesize $k$+$l$}}
\end{picture} 
\end{equation}
Taylor expanding the corresponding Feynman integral with respect to the 
momentum transfer $q$ around $q \f{=} 0$ generates additional loop momenta 
in the numerator, increases the power of one denominator, and reduces the 
original three-point function to a two-point function.
The resulting integrals in $D \f{=} 4 \f{-} 2\epsilon$
dimensions \cite{dimreg} are of the form 
\begin{equation}\label{integral0}
 \int\limits d^D k \; d^D l \;
 \frac{ (k\cdot p)^i (k\cdot q)^j
        (l\cdot p)^k (l\cdot q)^l
      }{P_{k+p,m_1}^{n} P_{k+l,m_2} P_{l,m_3}} 
\end{equation}
with integer $i$, $j$, $k$, $l$, $n$ and propagators 
$P_{k,m} \f{=} k^2 \f{-} m^2 \f{+} i\eta$. 
After a tensor decomposition with respect to $q$ and expressing the
remaining scalar products of $k$, $l$ and $p$ in terms of 
$P_{k+p,m_1}$, $P_{k+l,m_2}$, $P_{l,m_3}$ and the Mandelstam-type variables
\begin{eqnarray}
s_{12} &=& (p-l)^2  \\
s_{23} &=&     k^2 ,
\end{eqnarray}
(\ref{integral0}) can be reduced to a combination of the following basic 
sunset integrals 
\begin{equation}\label{eq:Tdef}
  T_{\alpha,\beta,n_1,n_2,n_3}\left(m_1^2,m_2^2,m_3^2;p^2\right) 
  \;\:\: = \;\:\:
  \pi^{-D} \int\limits d^D k \; d^D l \;
  \frac{s_{12}^{\alpha}\: s_{23}^{\beta}}
  {P_{k+p,m_1}^{n_1} P_{k+l,m_2}^{n_2} P_{l,m_3}^{n_3}} 
\end{equation}
and more simple graphs where one of the propagators has been cancelled. The
latter are essentially products of well known one-loop vacuum integrals
\cite{scalaroneloop} and thus easy to calculate, whereas the former are not
known analytically except for special combinations of masses. The
calculation of the quadratic and quartic charge-radii of the light 
pseudoscalar mesons in SU(3) chiral perturbation theory 
to the order ${\cal O}(p^6)$, for instance, requires the general 
nonzero mass case of $T_{\alpha,\beta,n_1,1,1}$ 
with $\alpha \f{+} \beta \f{\le} 7$ and $n_1 \f{\le} 6$ 
and illustrates that rather high tensors and powers of denominators
actually occur in practice \cite{chiPT}.

Sunset diagrams without numerators ($\alpha \f{=} \beta \f{=} 0$) 
have been studied in several papers. 
In \cite{BT-C} they were reduced, by direct integration in
momentum space \cite{Kreimer-CKK}, to two-dimensional integrals, which were 
then evaluated numerically. A one-dimensional integral representation, 
derived from a Feynman parametrization, was used in \cite{Ghinculov}. 
Another one-dimensional integral representation was obtained
from a dispersion relation in \cite{BBBB(W),Lunev}. 
For the case of equal masses, series expansions in $p^2$ were considered
in \cite{Kershaw-Mendels,BFT}, and for the general mass case, various
multiple series in $p^2$ and the masses $m_i^2$ were given in
\cite{Lunev,BBBS}. In \cite{ScharfDoc}, it was shown that the sunset 
diagram is a solution of a sixth order ordinary differential equation
in $p^2$.

Less is known about the above integrals $T_{\alpha,\beta,n_1,n_2,n_3}$
for positive $\alpha$, $\beta$. It is possible to regard them as master
integrals (i.e. two-point functions with five propagators) where two of the
propagators are massless and are raised to negative powers. In this way,
expansions in $p^2$ and $1/p^2$ can be calculated by the recursive
procedures described in \cite{D(S)T}. However, as one goes to higher orders
in the expansions, the complexity of the coefficients increases very
rapidly. Therefore, in practice, the number of terms that can be obtained
is limited, and may or may not be sufficient, depending on the accuracy
required and the value of $p^2$ considered.

In this paper we present a hybrid method for calculating the integrals
$T_{\alpha,\beta,n_1,n_2,n_3}$ with arbitrary nonzero masses. First, we
split off a Taylor polynomial in $p^2$, which enables us to calculate the
ultraviolet divergent part in an exact analytical form. Then, using a
dispersion relation, the remaining part, which is finite as $D \f{\to} 4$, is
expressed as a one-dimensional integral which can be evaluated numerically
to any desired precision. It is a generalization of the integral
representation for $T_{0,0,1,1,1}$ given in \cite{BBBB(W),Lunev}.

Our paper is divided into three parts: in section \ref{Kapitel2} the method
is described in some detail, in section \ref{Kapitel3} the structure of the
one-dimensional integration is analyzed with respect to numerical
instabilities, and in section \ref{Kapitel4} relations among the
$T_{\alpha,\beta,n_1,n_2,n_3}$ are established, which can be used to
minimize the number of evaluations of these integrals required in the
calculation of a physical process.


\section{Presentation of the method} 
\label{Kapitel2}

An important property of diagrams of the sunset topology is that
their ultraviolet divergences (both the overall divergence, and those
due to divergent subdiagrams) are polynomials in the external momentum 
$p^2$. Thus, they can easily be isolated by performing a Taylor expansion 
in this variable. Because we consider only diagrams with massive
propagators, infrared divergences cannot occur.

In the following discussion, the integrals (\ref{eq:Tdef}) will be written
as $T(p^2)$ for simplicity. They can be continued analytically into the
complex $p^2$-plane, with a branch cut along the positive real axis from
$p^2 \f{=} (m_1 \f{+} m_2 \f{+} m_3)^2$ to infinity. 
For any $p^2$ not on this cut, we may apply Cauchy's theorem, keeping 
$D \f{\neq} 4$ to regulate 
possible ultraviolet divergences, 
\begin{equation}
\label{eq:Cauchy1}
(1-{\cal T}^{(r)}) T(p^2) \;\:\:=\;\:\: 
 \frac{1}{2 \pi i} 
\int\limits_C d \zeta \; \frac{(p^2)^r}{(\zeta - p^2) \zeta^r} T(\zeta)
\end{equation}
where the operator ${\cal T}$ extracts the first $r$ terms of
the Taylor expansion in $p^2$,
\begin{equation}
{\cal T}^{(r)} \:\:\;=\:\:\;
\sum_{j=0}^{r-1}
\left. \frac{{(p^2)}^j}{j!}
\frac{\partial^j}{(\partial p^2)^j}
\right|_{p^2=0} \; ,
\end{equation}
and $C$ is any contour in the complex plane that encloses
the points $\zeta \f{=} 0$ 
and $\zeta \f{=} p^2$ 
and avoids the cut.
On the cut, $T(p^2)$ can be recovered from (\ref{eq:Cauchy1})
by letting $p^2$ approach the real $p^2$-axis from above. 

For $C$ we choose a path consisting of two straight lines, one along each
side of the cut, connected by an infinitesimally small circle around the
branch point $\zeta \f{=} (m_1 \f{+} m_2 \f{+} m_3)^2$, 
and a large circle with radius $R$.
If the number of subtractions $r$ is sufficiently large (this will be
specified more precisely later), the contribution from the large circle
will vanish in the limit 
$R \f{\to} \infty$, yielding 
\begin{equation}
\label{eq:Cauchy2}
(1-{\cal T}^{(r)}) T(p^2) \:\:\;=\:\:\; 
 \frac{1}{2 \pi i} 
 \int\limits_{(m_1+m_2+m_3)^2}^{\infty} \!\!\! d \zeta \;
 \frac{(p^2)^r}{(\zeta - p^2) \zeta^r} \{ T(\zeta+i0)-T(\zeta-i0) \}
 \; .
\end{equation}
At this stage, the left and right hand sides of eq.~(\ref{eq:Cauchy2}) are
still dimensionally regularized. However, the quantity
$\{T(\zeta \f{+} i0) \f{-} T(\zeta \f{-} i0) \} \f{=} 2 i Im T(\zeta)$ 
is finite in 4 dimensions, and therefore, if $r$ is chosen large enough,
the integral on the right hand side may be evaluated with $D \f{=} 4$. Thus
we decompose $T(p^2)$ into 
\begin{equation}
\label{eq:decomp}
 T(p^2)  \:\:\;=\:\:\;  T^A(p^2) + T^N(p^2) \; ,
\end{equation}
where
\begin{equation}
\label{eq:TA}
T^A(p^2) \:\:\;=\:\:\; {\cal T}^{(r)} T(p^2) 
\end{equation}
contains all the ultraviolet
divergences and will be calculated analytically, and 
\begin{equation}
\label{eq:dispersion}
T^N(p^2) \:\:\;=\:\:\;
\frac{1}{\pi} \int\limits_{(m_1+m_2+m_3)^2}^{\infty} \!\!\! d \zeta \;
\frac{(p^2)^r}{(\zeta - p^2) \zeta^r} Im T(\zeta) 
\end{equation}
is finite in $D \f{=} 4$ 
and will be calculated numerically.

In order to calculate the analytic part (\ref{eq:TA}), we express the
derivatives of $T(p^2)$ at 
$p^2 \f{=} 0$ 
in terms of derivatives with respect to
a parameter $\rho$, by which we multiply the momentum $p$: 
\begin{equation}
\frac{{(p^2)}^j}{j!} 
\left.
\frac{\partial^j T(p^2)}{(\partial p^2)^j} \right|_{p^2=0}
\:\:\;=\:\:\;
\frac{1}{(2j)!}
\left.
\frac{\partial^{2j} T({(\rho p)}^2)} {\partial \rho^{2j}}
\right|_{\rho=0} .
\end{equation}
These can be evaluated by differentiation underneath the integrals in
(\ref{eq:Tdef}). When $\rho$ is subsequently set equal to zero, the
external momentum $p$ disappears from the propagators, leaving a sum of
vacuum integrals of the form 
\begin{equation}
\label{eq:vacs}
 \pi^{-D} \int\limits d^D k \; d^D l \;
 \frac{{(k\cdot p)}^{\alpha'} {(l\cdot p)}^{\beta'}}
  { {(k^2-m_1^2)}^{n_1'} 
    {((k+l)^2-m_2^2)}^{n_2'} 
    {(l^2-m_3^2)}^{n_3'}} .
\end{equation}
Such integrals can be calculated recursively for arbitrary integers
$\alpha^\prime$, $\beta^\prime$, $n_i^\prime$ \cite{vacs1,vacs2,D(S)T}. 

Let us now consider the dispersion integral (\ref{eq:dispersion}) in the 
special case $n_1 \f{=} n_2 \f{=} n_3 \f{=} 1$. 
More general cases will be obtained later by differentiation with respect
to the masses. To determine the minimal number of subtractions $r$
required, we need some information about the behaviour of $T(p^2)$ as
$|p^2| \f{\to} \infty$, which can be obtained from general theorems on
asymptotic expansions of Feynman diagrams \cite{as-ex}. For our purposes,
it is enough to know that, in $D \f{=} 4 \f{-} 2\epsilon$ dimensions, 
the large $p^2$ expansion of $T_{\alpha,\beta,1,1,1}$ has the following
structure:
\begin{equation}
T_{\alpha,\beta,1,1,1}(p^2) =
  (-p^2)^{\alpha+\beta+1-2\epsilon}
    \sum_{\sigma=0}^{\infty} \frac{a_{\sigma}}{(p^2)^{\sigma}}
+ (-p^2)^{\alpha+\beta-\epsilon}
    \sum_{\sigma=0}^{\infty} \frac{b_{\sigma}}{(p^2)^{\sigma}}
+ (-p^2)^{\alpha+\beta-1}
    \sum_{\sigma=0}^{\infty} \frac{c_{\sigma}}{(p^2)^{\sigma}}
\; .
\end{equation}
This shows that, for $\epsilon$ close to zero, the contribution from the
large circle to (\ref{eq:Cauchy1}) will vanish in the limit 
$R \f{\to} \infty$ if we choose 
$r \f{\geq} \alpha \f{+} \beta \f{+} 2$. 
Although the calculations are simplest when $r$ is chosen minimally
$(r \f{=} \alpha \f{+} \beta \f{+} 2)$, 
we are free to use a larger value, which amounts to
shifting an additional finite contribution from $T^{N}$ to $T^{A}$. This
provides us with an important consistency check on the method. 

An application of Cutkosky's rules \cite{CutkoskyRules} gives the imaginary
part of $T_{\alpha,\beta,1,1,1}$ as a two dimensional integral over three
body phase space (the Dalitz plot): 
\begin{equation}
\label{eq:impart}
Im T_{\alpha,\beta,1,1,1}(p^2) 
   \:\:\;=\:\:\; 
   \frac{\pi}{p^2} \int\limits_{\Omega(p^2)} d s_{12} d s_{23}
   s_{12}^{\alpha} s_{23}^{\beta} \; ,
\end{equation}
where the integration region $\Omega(p^2)$ is given by
\begin{eqnarray}
\label{eq:region_det}
   4 m_1^2 m_2^2 m_3^2
 + (s_{12}-m_1^2-m_2^2)(s_{23}-m_2^2-m_3^2)
   (s_{13}-m_1^2-m_3^2)
&& \nonumber \\ - m_3^2 {(s_{12}-m_1^2-m_2^2)}^2
 - m_1^2 {(s_{23}-m_2^2-m_3^2)}^2
 - m_2^2 {(s_{13}-m_1^2-m_3^2)}^2
 & > & 0 
\end{eqnarray}
with
\begin{equation}
\label{eq:region_branch}
s_{12} > {(m_1+m_2)}^2 \, , \; \; \;
s_{23} > {(m_2+m_3)}^2 \, , \; \; \;
s_{13} > {(m_1+m_3)}^2 \, .
\end{equation}
In (\ref{eq:region_det})-(\ref{eq:region_branch}), 
$s_{13} \f{=} (k \f{+} p \f{+} l)^2$
depends on $s_{12}$ and $s_{23}$ as follows:
\begin{equation}
s_{13} = p^2 + m_1^2 + m_2^2 + m_3^2 - s_{12} - s_{23} \, .
\end{equation}

One of the integrations in (\ref{eq:impart}) can be performed easily;
e.g., at fixed $s_{23}$, the range of $s_{12}$ is
$A \f{-} B \f{<} s_{12} \f{<} A \f{+} B$ with
\begin{eqnarray}
  A&=&\frac{1}{2s_{23}}\Big(p^2(m_2^2-m_3^2+s_{23})-s_{23}^2+
  s_{23}(m_1^2+m_2^2+m_3^2)-m_1^2(m_2^2-m_3^2)\Big)
  \\
  B&=&\frac{1}{2s_{23}}\sqrt{\lambda(s_{23},p^2,m_1^2)\,
  \lambda(s_{23},m_2^2,m_3^2)} \:,
\end{eqnarray}
where 
$\lambda(x,y,z) \f{=} (x \f{-} y \f{-} z)^2 \f{-} 4yz$ 
is the K{\"a}ll{\'e}n function,
so that
\begin{equation}
\label{ABPzetaBezeichnungen}
  Im T_{\alpha,\beta,1,1,1}(p^2) = \frac{\pi}{p^2}\; 
  \theta(p^2-(m_1+m_2+m_3)^2) \!\!
  \int\limits_{(m_2+m_3)^2}^{(p-m_1)^2} \!\!
  ds_{23}\;\frac{2 s_{23}^\beta}{\alpha+1}\:\sum_{j=0}^{\left[\frac{\alpha}{2}
  \right]}\left({\alpha+1}\atop {2j+1}\right) A^{\alpha-2j}B^{2j+1} \:.
\end{equation}  
In general, performing the second integration gives complete elliptic
integrals \cite{BBBB(W)}. However, it is not necessary to evaluate them
explicitly here. Instead, we can insert the result
(\ref{ABPzetaBezeichnungen}) into the dispersion integral
(\ref{eq:dispersion}) and interchange the $d\zeta$ and $d s_{23}$
integrations: 
\begin{eqnarray}
\label{eq:ds23dz}
  (1-{\cal T}^{(r)}) \, T_{\alpha,\beta,1,1,1}(p^2)&=&
  \int\limits_{(m_2+m_3)^2}^\infty ds_{23}\; s_{23}^{\beta-1}
  \sqrt{\lambda(s_{23},m_2^2,m_3^2)}
  \label{zetaIntegral}
  \nonumber
  \\
  &&\quad\cdot\int\limits_{(\sqrt{s_{23}}+m_1)^2}^\infty
  d\zeta\:\frac{{(p^2)}^r}{(\zeta-p^2)\,\zeta^{r}} 
  \frac{\sqrt{\lambda(\zeta,m_1^2,s_{23})}}{\zeta}P(\zeta) \:,
\end{eqnarray}
where
\begin{equation}  
  P(\zeta) \;\:\:=\:\:\;
  \frac{1}{\alpha+1}\sum_{j=0}^{\left[\frac{\alpha}{2}\right]}
  \left({\alpha+1}\atop{2j+1}\right) A^{\alpha-2j}B^{2j} \Big|_{p^2=\zeta}
\end{equation}
is a polynomial in $\zeta$ of degree $\alpha$.
The $\zeta$-integrals in (\ref{eq:ds23dz}) can be recognized as
subtracted dispersion integral representations of the one-loop scalar
two-point function, 
\begin{eqnarray}
 B(m_1^2,m_2^2;p^2)
 &=&-i\pi^{-D/2}\int d^D\!k\;\frac{1}{((k+p)^2-m_1^2)(k^2-m_2^2)}
 \:,
\end{eqnarray}
in which one of the masses is replaced with $s_{23}$:
\begin{eqnarray}
\label{AbfallimUnendl}
  (1-{\cal T}^{(k)}) \, B(m_1^2,s_{23};p^2)&=&
  \int\limits_{(\sqrt{s_{23}}+m_1)^2}^\infty
  d\zeta\:\frac{{(p^2)}^k}{(\zeta-p^2)\,\zeta^{k}}\:
  \frac{\sqrt{\lambda(\zeta,m_1^2,s_{23})}}{\zeta} \:.
\end{eqnarray}

Thus, the numerical part $T^N_{\alpha,\beta,1,1,1}(p^2)$ is given by a sum
of one-dimensional integrals of the form
\begin{equation}
\label{eq:TN1dim}
  \int\limits_{(m_2+m_3)^2}^\infty ds_{23}\; s_{23}^l
  \sqrt{\lambda(s_{23},m_2^2,m_3^2)}
  \, \left(1-{\cal T}^{(k)}\right) B(m_1^2,s_{23};p^2)
  \;,
\end{equation}
with
\begin{eqnarray}
  k &=& \mbox{$r \f{-} \alpha$, $r \f{-} \alpha \f{+} 1$, $\f{\ldots}$ , $r$} 
  \\
  l &=& \mbox{$\beta \f{-} \alpha \f{-} 1$, $\beta \f{-} \alpha$,
   $\f{\ldots}$ , $\beta \f{+} \alpha \f{-} 1 \f{+} k \f{-} r$.}
\end{eqnarray}
We use the following well known expressions for $B(m_1^2,s_{23};p^2)$
\cite{scalaroneloop}:
\begin{eqnarray}
   B(m_1^2,s_{23};p^2)
  &=& \frac{1}{\epsilon}-\gamma-\int_0^1 dx\log(m_1^2x+s_{23}(1-x)
  -p^2 x(1-x))
  \label{FeynmanParameter}
  \\
  &=& \frac{1}{\epsilon}-\gamma+2-\log m_1^2 + F(m_1^2;s_{23};p^2) \:,
  \label{oneloopfunction}
\end{eqnarray}
where 
\begin{eqnarray}
  F(m_1^2;s_{23};p^2) &=& x_1 \log\left(1-\frac{1}{x_1}\right) 
  +  x_2 \log\left(1-\frac{1}{x_2}\right) \, ,
  \label{nichttrAnteil}
\end{eqnarray}
$x_{1,2}$ are the zeros of the argument of the logarithm in the
Feynman parameter integral,
\begin{equation}
x_{1,2} \:\:\;=\:\:\; \frac{1}{2 p^2} \left\{ p^2 + s_{23} - m_1^2 \pm
          \sqrt{ \lambda(p^2,s_{23}, m_1^2) } \right\} \, ,
\end{equation}
and $\gamma$ is Euler's constant. The derivatives at $p^2 \f{=} 0$ 
that enter (\ref{eq:TN1dim}) can easily be obtained from
(\ref{FeynmanParameter}), e.g.: 
\begin{eqnarray}
  \label{pAbleitung0}
  B(m_1^2,s_{23};0)&=& \frac{1}{\epsilon}-\gamma+1-\log m_1^2
  + \frac{s_{23}}{m_1^2-s_{23}}
  \log\frac{s_{23}}{m_1^2}
  \\
  \label{pAbleitung1}
  \frac{\partial B}{\partial p^2}(m_1^2,s_{23};0) &=&
  \frac{m_1^2+s_{23}}{2(m_1^2-s_{23})^2}
  +\frac{m_1^2 s_{23}}{(m_1^2-s_{23})^3}\log\frac{s_{23}}{m_1^2} \:.
\end{eqnarray}

Integrals $T_{\alpha,\beta,n_1,n_2,n_3}$ where one or more of
the indices $n_1$, $n_2$, $n_3$ are greater than one are related to 
$T_{\alpha,\beta,1,1,1}$ by differentiation with respect to $m_1^2$, 
$m_2^2$, $m_3^2$:
\begin{eqnarray}
\lefteqn{
T_{\alpha,\beta,n_1+1,n_2+1,n_3+1} \left(m_1^2,m_2^2,m_3^2;p^2\right)
} && \nonumber \\
&=&
\frac{1}{n_1! n_2! n_3!} \;
\frac{\partial^{n_1+n_2+n_3}}
{(\partial m_1^2)^{n_1} (\partial m_2^2)^{n_2} 
 (\partial m_3^2)^{n_3}} \;
 T_{\alpha,\beta,1,1,1}\left(m_1^2,m_2^2,m_3^2;p^2\right) \:.
\end{eqnarray}
Differentiating $T^A_{\alpha,\beta,1,1,1}$ in this manner leads to vacuum
integrals of the same kind as in (\ref{eq:vacs}). For the numerical part,
$T^N_{\alpha,\beta,n_1,n_2,n_3}$, it is necessary to differentiate the
integrals (\ref{eq:TN1dim}). This can be done underneath the integral sign,
provided one first shifts the integration variable
$s_{23} \f{\to} s'_{23} \f{+} (m_2 \f{+} m_3)^2$,
in order to avoid difficulties
due to the square root like behaviour of the integrand at the lower
end point. Furthermore, the mass derivatives of the one-loop function $B$ are
calculated most effectively using the fact that the nontrivial part $F$
essentially reproduces itself:
\begin{equation}
 \frac{\partial B(m_1^2,s_{23};p^2)}{\partial m_1^2} = 
 \frac{1}{\lambda(m_1^2,s_{23},p^2)} \left[
 (m_1^2-s_{23}-p^2) F(m_1^2;s_{23};p^2) 
 + 2s_{23} \log\frac{m_1^2}{s_{23}} \right] \:.
\end{equation}


\section{Numerical analysis}
\label{Kapitel3}

When we are below the physical threshold, i.e. 
$p^2 \f{<} (m_1 \f{+} m_2 \f{+} m_3)^2$,
the integrands in (\ref{eq:TN1dim}) are smooth and their numerical
integration does not present any problems. However, some care
is needed to evaluate the integrands with sufficient accuracy. The 
difficulties come from the following regions:
\begin{description}
\item[\boldmath\bf $s_{23} \f{\to} \infty$:]   
The expansion of $B(m_1^2,s_{23};p^2)$ in $1/s_{23}$ reads
  (see also \cite{BBBS,BBBB(W)}):
\begin{equation}
\label{b0-asex}
B(m_1^2,s;p^2) 
 = \frac{1}{\epsilon} - \gamma + 1 - \log m_1^2
 + \sum_{\sigma=1}^{\infty} \frac{1}{s_{23}^{\sigma}}
   \sum_{k=0}^{\sigma} (p^2)^k (m_1^2)^{\sigma-k}
   \left[ a_{k \sigma} \log \frac{m_1^2}{s_{23}}
        + b_{k \sigma} \right] \; ,
\end{equation}
where
\begin{eqnarray}
a_{k \sigma} &=& \frac{ \sigma! (\sigma-1)! }
{ k! (k+1)! (\sigma-k)! (\sigma-k-1)! } 
 \quad \mbox{for $k \f{=} 0, \f{\ldots} , \sigma \f{-} 1$}
\nonumber \\
a_{\sigma \sigma} &=& 0
\nonumber \\
b_{0 \sigma} &=& 0
\nonumber \\
b_{k \sigma} &=&
a_{k \sigma}
\{ \psi(\sigma) + \psi(\sigma+1)
  -\psi(\sigma-k) - \psi(\sigma-k+1) \}
\nonumber\\
&& \mbox{for $k \f{=} 1, \f{\ldots} , \sigma \f{-} 1$ and
   $\psi(n) \f{=} -\gamma \f{+} \frac{1}{1} \f{+} \frac{1}{2}
   \f{+} \f{\ldots} \f{+} \frac{1}{n-1}$ }  
\nonumber\\
b_{\sigma \sigma} &=& \frac{1}{\sigma(\sigma+1)} \:.
\end{eqnarray}

 By omitting all summands containing powers of $p^2$ less than $r$, it
 is seen that the leading term of 
 $(1 \f{-}{\cal T}^{(r)}) B(p^2)$ is of order $1/s_{23}^r$. 
 Consequently, using the exact 
 formulae gives rise to large cancellations, and it is much better to
 replace the integrand by its asymptotic expansion whenever
 $s_{23} \f{\geq} S\,(m_1 \f{+} \sqrt{p^2})^2$, 
 $S \f{=} \mbox{\em const}$. 
 For example, we obtained numerically stable results
 with the choice $S=4$ when we truncated the series (\ref{b0-asex}) at
 $\sigma=50$.
\item[\boldmath\bf$s_{23} \f{\approx} m_1^2$:]
 If $m_2 \f{+} m_3 \f{<} m_1$, 
 the region $s_{23} \f{\approx} m_1^2$ lies within the integration
 interval and causes problems due to factors of 
 $(m_1^2 \f{-} s_{23})$ in the 
 denominator (see e.g. (\ref{pAbleitung0}) f). Nevertheless, the singularities
 of the different parts exactly cancel: the integrand is smooth at
 $s_{23} \f{\approx} m_1^2$ 
 and can be approximated by a Taylor-polynomial or 
 rather a Pad{\'e}-approximant. For integrals $T_{\alpha,\beta,n_1,1,1}$ with 
 $\alpha \f{+} \beta \f{\le} 7$ and 
 $n_1 \f{\le} 6$ 
 a diagonal Pad{\'e}-approximation of 
 order $(20,20)$ in $s_{23} \f{\in} [0.2m_1^2,5m_1^2]$ 
 turned out to be a suitable choice.
\item[\boldmath\bf$s_{23} \f{\approx} (m_1 \f{\pm} \sqrt{p^2})^2$:]
 Differentiation of $B(m_1^2,s_{23};p^2)$ with respect to $m_1^2$ 
 (or $s_{23}$) brings down a factor $\lambda(m_1^2,s_{23},p^2)$
 in the denominator, which vanishes for
 $s_{23} \f{=} (m_1 \f{\pm}\! \sqrt{p^2})^2$. 
 Since $B$ and its derivatives are smooth 
 at $s_{23} \f{=} (m_1 \f{+} \sqrt{p^2})^2$ 
 and (below the physical threshold)
 at $s_{23} \f{=} (m_1 \f{-} \sqrt{p^2})^2$, 
 a simple Taylor-approximation (up to order 10, e.g.) remedies the
 instability. 
\end{description}
As a rule, cancellations are more severe when higher mass derivatives
are taken or more terms of the Taylor series need to be subtracted.

Above the threshold, $p^2 \f{>} (m_1 \f{+} m_2 \f{+} m_3)^2$, 
singularities appear in
the integrand at $s_{23} \f{=} (\sqrt{p^2} \f{-} m_1)^2$, 
where the threshold of
$B(m_1^2,s_{23};p^2)$ is crossed. Although $B$ itself is still finite at this 
point, its derivatives are not. Therefore, noting that $B(m_1^2,s_{23};p^2)$
is an analytic function of $s_{23}$ in the lower half of the complex
plane, we deform the integration contour so as to steer clear of the singular
point. The simplest possibility is to choose a straight line,
\begin{equation}
  s_{23} \;=\; 
  (m_2+m_3)^2+e^{-\mbox{i}\theta}s' \, , \; \; \; \;  0<s'<\infty \; ,
\end{equation}
at some angle $0 \f{<} \theta \f{<} \pi/2$ 
with respect to the real $s_{23}$-axis.
Along the deformed contour the integrand is smooth and can be integrated
easily\footnote{This approach was also used in \cite{Ghinculov}, where
similar problems were encountered.}. 
The fact that the result should not depend on $\theta$ can be used as a
check. Notice, that such a rotation of the contour also avoids the problems
at $s_{23} \f{\approx} m_1^2$ and at $s_{23} \f{\approx} (m_1 \f{\pm}
\sqrt{p^2})^2$ mentioned above, unless the left end of the integration
contour happens to be a critical point itself. 

We compared a number of special cases with results available in the
literature. We computed the first 17 Taylor coefficients of
$T_{0,0,1,1,1}(m_1^2,m_2^2,m_3^2;p^2)$ and verified that they satisfy
the five-term recurrence relation derived in \cite{ScharfDoc}.
Another check involved the three point function (\ref{fig3pt}), which can be
calculated analytically in the special case $p^2=(p+q)^2=m_1^2=m_2^2=m_3^2$
\cite{Bessis}. We expanded the result around $q^2=0$ and found agreement
with the first 3 terms as calculated by the algorithm described in this
paper. We also checked that our algorithm reproduces the $\epsilon^{-2}$,
$\epsilon^{-1}$ and $\epsilon^0$ terms of
$T_{0,0,1,1,1}(m^2,m^2,m^2;p^2=m^2)$, which can be found in \cite{BFT}. 

Finally, as an example, we would like to give some concrete results for 
$T_{0,3,4,1,1}$. We take the masses $m_1 \f{=} 0.28$,
$m_2 \f{=} 1$, $m_3 \f{=} 1.143328474$, corresponding to the masses of the
$\pi$, $K$ and $\eta_8$ mesons in chiral perturbation theory (in units of
$m_K$), and two values of $p^2$, one below ($p^2 \f{=} 1$) and one
above ($p^2 \f{=} 9$) the threshold. The parameter $r$ denotes the number
of Taylor terms subtracted from $T_{0,3,4,1,1}$ and 
defines $T_{0,3,4,1,1}^A \f{=} {\cal T}^{(r)} T_{0,3,4,1,1}$, 
cf. (\ref{eq:TA}).
The divergent part of $T_{0,3,4,1,1}^A$ is given by
\begin{eqnarray} 
T_{0,3,4,1,1}^{A,div} &=& 
  - \frac{1}{2 \epsilon^2} \left( 6 p^2 + 4 m_1^2  + m_2^2 + m_3^2 \right) 
  + \frac{1}{6 \epsilon} 
    \Big( 18 p^2 - 10 m_1^2 - 9 m_3^2 - 9 m_2^2     
\\
 && - \frac{(p^2)^3}{m_1^4} + 12 \frac{(p^2)^2}{m_1^2}  
    + 36 \gamma p^2 + 24 \gamma m_1^2 + 6 \gamma m_2^2 + 6 \gamma m_3^2
\nonumber\\
 && + 36 p^2 \log(m_1^2) + 24 m_1^2 \log(m_1^2) 
    + 6 m_2^2 \log(m_2^2) + 6 m_3^2 \log(m_3^2) \Big) \, .
\nonumber
\end{eqnarray}  
The finite (order $\epsilon^0$) part of $T_{0,3,4,1,1}^A$, and
$T_{0,3,4,1,1}^N$ are given in the following table for different values
of $r$:
\begin{center}
\begin{tabular}{|r||r|r||r|r|}
\hline \rule[-0.2ex]{0em}{2.6ex}
$r$ & $T_{0,3,4,1,1}^N(1)$ & $T_{0,3,4,1,1}^{A,fin}(1)$ 
    & \multicolumn{1}{c|}{$T_{0,3,4,1,1}^N(9)$} & $T_{0,3,4,1,1}^{A,fin}(9)$
\\ 
\hline\hline
5 & $-0.2879782058$ & $-43.6425974985$ 
  & $3267.7085 \f{-} 38384.2371\:i$ & $-61507.6741$
\\
\hline
6 & $-0.0318842307$ & $-43.8986914736$ 
  & $18389.8016 \f{-} 38384.2371\:i$ & $-76629.7672$
\\
\hline
7 & $-0.0040709644$ & $-43.9265047399$ 
  & $33170.9116 \f{-} 38384.2371\:i$ & $-91410.8772$
\\
\hline
8 & $-0.0005618259$ & $-43.9300138784$ 
  & $49955.0126 \f{-} 38384.2371\:i$ & $-108194.9782$ 
\\
\hline\hline \rule[-0.2ex]{0em}{2.6ex}
& \multicolumn{2}{c||}{$T_{0,3,4,1,1}^{fin}(1) \f{=} -43.9305757043$}  
& \multicolumn{2}{c|}{$T_{0,3,4,1,1}^{fin}(9) 
   \f{=} -58239.9656 \f{-} 38384.2371\:i$} 
\\
\hline
\end{tabular}
\end{center}
The bottom row shows the total finite part 
$T_{0,3,4,1,1}^{fin} \f{=} T_{0,3,4,1,1}^N \f{+} T_{0,3,4,1,1}^{A,fin}$,
which is independent of $r$, as it should be. Below the threshold, the
Taylor series converges, so that in cases where only moderate accuracy is
required (eg., 4 digits), it may be possible to neglect the numerical
contribution $T_{0,3,4,1,1}^N$. In our application \cite{chiPT}, however,
we sometimes needed as many as 8 digits, due to large cancellations
between different $T_{\alpha,\beta,n_1,n_2,n_3}$'s. In any case,
$T_{0,3,4,1,1}^N$ is indispensable above the threshold, where the Taylor 
series diverges.


\section{Relations among the basic sunset integrals}\label{Kapitel4}

The method presented in the previous section enables us to calculate
analytically the divergent part and (for given masses and external momentum) 
numerically the finite part of each $T_{\alpha,\beta,n_1,n_2,n_3}$.
Nevertheless, the question arises whether the set of
$T_{\alpha,\beta,n_1,n_2,n_3}$ can be reduced further, both for practical and
for theoretical reasons. In fact, there are three sources of relations,
\begin{itemize}
  \item[-] permutation symmetry, if some masses are equal 
   (e.g.\ $T_{\alpha,\beta,1,1,1} \f{=} T_{\beta,\alpha,1,1,1}$ 
    if $m_1 \f{=} m_3$),
  \item[-] integration by parts identities \cite{dimreg,vacs1,ibp},
  \item[-] subloop tensor decomposition, 
\end{itemize}
the third of which we would like to exploit in the sequel.

The most nontrivial part of a fixed 
$T_{\alpha,\beta,n_1,n_2,n_3}$ 
is the integral 
\begin{equation}
  \int d^D k \; d^D l \; 
  \frac{ (l \cdot p)^{\alpha} \: (k^2)^\beta }
       { P_{k+p,m_1}^{n_1} P_{k+l,m_2}^{n_2} 
       P_{l,m_3}^{n_3} }
  \;\:\: = \;\:\:
  \int d^D k \; \frac{(k^2)^{\beta}}{ P_{k+p,m_1}^{n_1} }
  \; p_{\mu_1} \ldots p_{\mu_{\alpha}}
  \; \int d^D l \; 
  \frac{ l^{\mu_1} \ldots l^{\mu_{\alpha}} }
       { P_{k+l,m_2}^{n_2} P_{l,m_3}^{n_3} } \: .  
\end{equation}
Consider for a moment just the $l$-integration and decompose its
tensor structure in terms of $g^{\mu\nu}$ and $k^\mu$, which is the 
external momentum of the $l$-subloop \cite{PasVelt}. This yields 
\begin{equation}
 (k^2)^{-\alpha} \int d^D l \; \frac{ N }{ P_{k+l,m_2}^{n_2} 
 P_{l,m_3}^{n_3} } \: ,
\end{equation}
for the $l$-integral, where $N$ is a homogeneous polynomial of degree
$2\alpha$ in $k^2$, $k\cdot l$, $l^2$, $k\cdot p$ and $p^2$. After
rewriting the whole integral 
\begin{equation}
  \int d^D k \; d^D l \; 
  \frac{ (k^2)^{\beta-\alpha} \: N }{ P_{k+p,m_1}^{n_1} 
  P_{k+l,m_2}^{n_2}  P_{l,m_3}^{n_3} }   
\end{equation}
in terms of $T$'s again we have reduced $T_{\alpha,\beta,n_1,n_2,n_3}$
to integrals $T_{\alpha^\prime,\beta^\prime,n_1^\prime,n_2^\prime,n_3^\prime}$
with $\alpha^\prime \f{=} 0$, 
$\beta^\prime \f{\in} \{\beta \f{-} \alpha, \f{\ldots} , \beta \f{+} \alpha\}$,
$n_1^\prime \f{=} n_1$, 
$n_2^\prime \f{=} n_2$, 
$n_3^\prime \f{=} n_3$ 
or
$n_1^\prime \f{+} n_2^\prime \f{+} n_3^\prime \f{<}  n_1 \f{+} n_2 \f{+} n_3$. 

This procedure can be illustrated by the following graphical picture.
For each triplet
$(n_1,n_2,n_3)$ identify the integrals $T_{\alpha,\beta,n_1,n_2,n_3}$
with the points of an associated two-dimen\-sional $(\alpha,\beta)$-lattice
and order the lattices according to $n_1 \f{+} n_2 \f{+} n_3$. Then the
point $(\alpha,\beta)$ in the $(n_1,n_2,n_3)$-lattice can be 'projected' to
the points 
$(0,\beta \f{-} \alpha), \f{\ldots} , (0,\beta \f{+} \alpha)$ 
on the $\beta$-axis and points in lower lying lattices. Analogously, by
changing the flow of the external momentum through the diagram, one can
project a point on the $\alpha$-axis. 

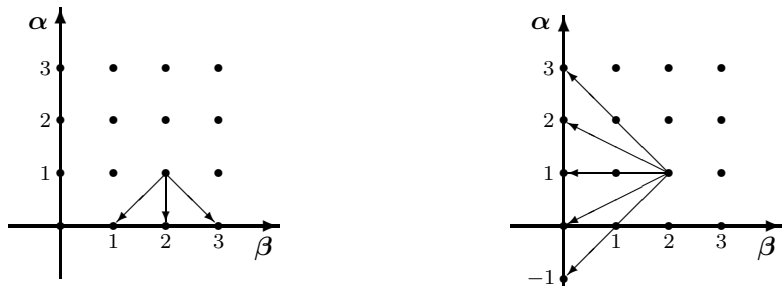
\begin{center}
\unitlength1mm
\scriptsize
\begin{picture}(35,35)
\thicklines
\put(1,8){\vector(1,0){36}}
\put(8,1){\vector(0,1){36}}
\put(8,8){\circle*{1}}
\put(15,8){\circle*{1}}
\put(22,8){\circle*{1}}
\put(29,8){\circle*{1}}
\put(8,15){\circle*{1}}
\put(15,15){\circle*{1}}
\put(22,15){\circle*{1}}
\put(29,15){\circle*{1}}
\put(8,22){\circle*{1}}
\put(15,22){\circle*{1}}
\put(22,22){\circle*{1}}
\put(29,22){\circle*{1}}
\put(8,29){\circle*{1}}
\put(15,29){\circle*{1}}
\put(22,29){\circle*{1}}
\put(29,29){\circle*{1}}
\put(15,6){\makebox(0,0){$1$}}
\put(22,6){\makebox(0,0){$2$}}
\put(29,6){\makebox(0,0){$3$}}
\put(35,5){\makebox(0,0){\footnotesize\boldmath\bf$\beta$}}
\put(6,15){\makebox(0,0){$1$}}
\put(6,22){\makebox(0,0){$2$}}
\put(6,29){\makebox(0,0){$3$}}
\put(5,35){\makebox(0,0){\footnotesize\boldmath\bf$\alpha$}}
\thinlines
\put(22,15){\vector(0,-1){6.5}}
\put(22,15){\vector(1,-1){6.5}}
\put(22,15){\vector(-1,-1){6.5}}
\end{picture}
\hspace{10em}
\begin{picture}(35,35)
\thicklines
\put(1,8){\vector(1,0){36}}
\put(8,0){\vector(0,1){36}}
\put(8,1){\circle*{1}}
\put(8,8){\circle*{1}}
\put(15,8){\circle*{1}}
\put(22,8){\circle*{1}}
\put(29,8){\circle*{1}}
\put(8,15){\circle*{1}}
\put(15,15){\circle*{1}}
\put(22,15){\circle*{1}}
\put(29,15){\circle*{1}}
\put(8,22){\circle*{1}}
\put(15,22){\circle*{1}}
\put(22,22){\circle*{1}}
\put(29,22){\circle*{1}}
\put(8,29){\circle*{1}}
\put(15,29){\circle*{1}}
\put(22,29){\circle*{1}}
\put(29,29){\circle*{1}}
\put(15,6){\makebox(0,0){$1$}}
\put(22,6){\makebox(0,0){$2$}}
\put(29,6){\makebox(0,0){$3$}}
\put(35,5){\makebox(0,0){\footnotesize\boldmath\bf$\beta$}}
\put(5,1){\makebox(0,0){$-1$}}
\put(6,15){\makebox(0,0){$1$}}
\put(6,22){\makebox(0,0){$2$}}
\put(6,29){\makebox(0,0){$3$}}
\put(5,35){\makebox(0,0){\footnotesize\boldmath\bf$\alpha$}}
\thinlines
\put(22,15){\vector(-1,1){13.5}}
\put(22,15){\vector(-2,1){13.5}}
\put(22,15){\vector(-1,0){13.5}}
\put(22,15){\vector(-2,-1){13.5}}
\put(22,15){\vector(-1,-1){13.5}}
\end{picture}
\end{center}
On the one hand, these projections show that any 
$T_{\alpha,\beta,n_1,n_2,n_3}$ can be expressed in terms of integrals on the
positive axes ($\alpha \f{=} 0$ 
or $\beta \f{=} 0$ 
and $\min(\alpha,\beta) \f{\ge} 0$)
and trivial one-loop-products, so that at most the positive axes are
independent. On the other hand, they entail further relations for fixed
$(n_1,n_2,n_3)$: 
\begin{itemize}
\item[-]
  First, there are two possibilities to project the points $(\alpha,\alpha)$ 
  on the diagonal onto a positive axis,
  which sets up a relation between the point $(2\alpha,0)$ and the points
  $(\alpha^\prime,0)$ with 
  $0 \f{\le} \alpha^\prime \f{<} 2\alpha$ and $(0,\beta)$ 
  with $0 \f{\le} \beta \f{\le} 2\alpha$. 
  In other words, the even points on one of the positive axes, e.g. the
  positive $\alpha$-axis, are eliminated from the minimal set of $T$'s. 
\item[-]
  Second, equating the two projections of the point 
  $(\alpha, \alpha \f{+} 1)$ in the 
  line below the diagonal (cf. the above figures)
  yields a relation between the point 
  $(2\alpha \f{+} 1,0)$ 
  and the points $(\alpha^\prime,0)$ with 
  $0 \f{\le} \alpha^\prime \f{\le} 2\alpha$,
  $(0,\beta)$ with $0 \f{\le} \beta \f{\le} 2\alpha \f{+} 1$
  and $(-1,0)$. The integral $(-1,0)$, which
  is of a different topology, can be expressed in terms of $(1,0)$, $(0,0)$ and
  $(0,1)$ by equating the two projections of $(0,1)$. In this way the odd
  points of the positive $\alpha$-axis with the exception of $(1,0)$ are
  eliminated from the minimal set of $T$'s.
\end{itemize}
The point $(1,0)$ can be eliminated as well by equating the two projections 
of $(2,1)$ and using the above results for $(3,0)$ and $(2,0)$.
But, unlike the relations for the points $(\alpha,0)$ with
$\alpha \f{>} 1$, the 
relation found in this way contains a denominator 
$(m_1^2 \f{-} p^2)(m_2^2 \f{-} m_3^2)$
and is therefore not applicable in all mass cases.
For this reason, we prefer to keep $T_{1,0,n_1,n_2,n_3}$ in our set of 
basic integrals, along with the integrals $T_{0,\beta,n_1,n_2,n_3}$ with 
$\beta \f{\geq} 0$.


\section{Conclusion}

In this paper we presented a method to calculate the dimensionally regularized
sunset integrals $T_{\alpha,\beta,n_1,n_2,n_3}(m_1^2,m_2^2,m_3^2;p^2)$, see
(\ref{eq:Tdef}), with arbitrary nonzero masses and arbitrary momentum $p^2$, 
both below and above the threshold. The simple topology of such
dia\-grams -- they can only be cut in one way -- allows their divergences
to be extracted by subtracting a finite number of terms of their
Taylor expansion in $p^2$ around $p^2 \f{=} 0$. The subtracted terms are
vacuum diagrams that are known analytically. The remainder, which is
finite in $D \f{=} 4$ dimensions, is transformed into a well-behaved
one-dimensional integral that can be evaluated numerically. Furthermore,
we discussed a set of linear relations between different
$T_{\alpha,\beta,n_1,n_2,n_3}$'s and suggested a reduction mechanism to a
basic set of these integrals, namely $T_{1,0,n_1,n_2,n_3}$ and 
$T_{0,\beta,n_1,n_2,n_3}$ with $\beta \f{\geq} 0$.

The algorithm was developed in order to meet the needs of a calculation
in SU(3) chiral perturbation theory, but it can be applied in any field
theory describing massive particles with four-point interaction vertices.
The basic idea of the method can be extended to any two-loop diagram
containing a subgraph which itself corresponds to a two-point function
\cite{BBBB(W)}; in particular, the full three-point graph (\ref{fig3pt})
can be handled in this way for arbitrary momentum transfer $q$. For chiral
perturbation theory the techniques presented in this paper are a step
towards taking into account the full mass dependence of the nontrivial
two-loop diagrams which occur at the order ${\cal O}(p^6)$ of the momentum
expansion. 

We thank D.J. Broadhurst for some valuable suggestions, J. Gasser for
drawing our attention to ref. \cite{Bessis}, and K. Schilcher for helpful
discussions. J.B.T. was supported by the Graduiertenkolleg
``Elementarteilchenphysik bei mittleren und hohen Energien'' in Mainz and
P.P. by the ``Studienstiftung des deutschen Volkes''.



\end{document}